\documentstyle[thmsa,sw20lart]{article}
\input{tcilatex}
\begin{document}

\title{The assumptions underlying the consistent-amplitude approach to quantum
theory}
\author{Ariel Caticha \\
%EndAName
{\small Department of Physics, University at Albany-SUNY, Albany, NY 12222%
\thanks{%
E-mail: ariel@cnsvax.albany.edu}}}
\date{}
\maketitle

\begin{abstract}
In this note I expand further on the main assumptions leading to the
consistent-amplitude approach to quantum theory and I offer a reply to Jerry
Finkelstein's recent comment (quant-ph/9809017) concerning my argument for
the linearity of quantum mechanics (Phys. Rev. {\bf A57}, 1572 (1998)).
\end{abstract}

The purpose of this note is twofold: it is intended as a reply to a recent
comment made by Jerry Finkelstein \cite{Finkelstein98} on my work on the
consistent-amplitude approach to quantum theory (CAQT) \cite{Caticha98a}\cite
{Caticha98b}\cite{Caticha98c} and, in the process, I will provide a brief
overview emphasizing the main assumptions behind the CAQT.

The objective of CAQT has been to justify the formalism of quantum theory in
the hope that this would not only clarify the formal connections among the
various postulates of quantum theory but also illuminate the issue of how
the formalism should be interpreted. In this respect the traditional
approach has been to first set up the formalism and then try to find out
what it all means. This problem of attributing a physical meaning to
mathematical constructs is a notoriously difficult one. A well known ancient
example is the theory of probability. There the subject of the mathematical
structure of the theory has been settled for a very long time but questions
about the actual interpretation -- what a probability actually means --
remain to this day very controversial. So, rather than take the standard
quantum theory as axiomatized, say by von Neumann, and then, append to it an
interpretation, the approach I have taken has been to try to build the
formalism and its interpretation simultaneously. Thus, the hope is that by
the time the formalism is completed, at least some important
interpretational issues will have been settled.

I will address Finkelstein's comment in the context of a very brief summary:
CAQT is formulated as the only consistent way to manipulate the amplitudes
for quantum processes; the result is the standard quantum theory. We proceed
in several steps; effectively, each step represents an assumption. The first
and most crucial assumption is a decision about the subject matter. What
problem is quantum mechanics trying to solve? We choose a pragmatic,
operational approach: statements about a system are identified with those
experimental setups designed to test them. Quoting from \cite{Caticha98b}:
``Our goal is to predict the outcomes of experiments and the strategy is to
establish a network of relations among setups in the hope that information
about some setups might be helpful in making predictions about others.'' We
find that there are two basic kinds of relations among setups, which we call 
$and$ and $or$. These relations or operations represent our idealized
ability to build more complex setups out of simpler ones, either by placing
them in ``series'' or in ``parallel''. The identification of the $and/or$
relations, as well as their properties (associativity, distributivity, etc.)
is crucial to establishing the goal and the subject of quantum mechanics.
The first assumption is

\begin{itemize}
\item[{\bf A1.}]  The goal of quantum theory is to predict the outcomes of
experiments involving setups built from parts connected through $and$ and $or
$.
\end{itemize}

\noindent It is important to emphasize that this{\em \ }quantum theory
coincides with the standard Copenhagen quantum theory (see \cite{Stapp72}).
The contribution, at this point, has been to make explicit the relations $%
and/or$ which are normally implicit in the Feynman approach.

The next step involves an assumption about the means for handling these
relations $and/or$ quantitatively:

\begin{itemize}
\item[{\bf A2.}]  We seek a mathematical representation of $and/or$ by
assigning to each setup $a$ a {\em single} complex number $\phi (a)$ in such
a way that relations among setups translate into relations among the
corresponding complex numbers. The requirement that a single complex number
be assigned to a setup is a consistency constraint: if there are two
different ways to compute $\phi (a)$ the two answers must agree.
\end{itemize}

\noindent Quoting from \cite{Caticha98a}: ``There is no reason why such a
representation should exist, but if it does exist, and this is, perhaps,
just another instance of the ``unusual effectiveness of mathematics in the
physical sciences,'' then there is no reason not to take advantage of it.''
And again, quoting from \cite{Caticha98b}: ``Why should such a
representation exist? It need not, but all of physics consists of
representing elements of reality, or relations among these elements, or our
information about them, by mathematical objects. The existence of such
representations may be mysterious, but it is not surprising; there are too
many examples.'' The remarkable consequence of the existence of one
representation is the theorem that it is always possible to switch to
another one equivalent to it and considerably more convenient in which $and$
and $or$ are respectively represented by multiplication and addition. This
is the main result of \cite{Caticha98a}. 

In \cite{Caticha98a}, anticipating results to be obtained in the later
sections of \cite{Caticha98b}, such as for example, the fact that the time
evolution of these complex numbers is given by a linear Schr\"{o}dinger-like
equation and that their modulus squared yields probabilities, I wrote
``Complex numbers assigned in this way are called `amplitudes'''.%
\index{amplitudes} In retrospect perhaps this was pedagogically misleading
because it suggests that refs. \cite{Caticha98a}\cite{Caticha98b} can be
summarized as follows: If one attributes to quantum mechanical amplitudes,
not only all those properties to which we have become accustomed, but also
the additional requirement that they provide a representation for $and/or$
then one finds the Schr\"{o}dinger equation must be linear or else the
theory is inconsistent. Put in this way it is reasonable to conclude as does
Finkelstein \cite{Finkelstein98}: ``...Thus it is the demand that the
amplitudes form a `representation' with which nonlinear theories are
inconsistent.'' But one should not put it this way: I do not take the
amplitudes we all know and append an {\em additional} {\em requirement}, I
take complex numbers which carry no previous burden of preconceived meanings
and are totally arbitrary except for the {\em only requirement} that they
provide a representation of $and/or$. To avoid further misconceptions, in
the rest of this note, I will refer to these complex numbers after they have
been suitably regraded so that the sum and product rules hold, by $CN$s.
Toward the end I will comment on the implications of the interesting logical
possibility that such a representation of $and/or$ might not exist.

Thus, $CN$s are tools for reasoning that encode information about how one
builds complicated setups by combining more elementary ones. Incidentally,
as a consequence of the sum and product rules, the time evolution of these $%
CN$s is given by a linear equation, and this allows the introduction of a `$%
CN$-function' (which I will eventually call the `wave function') as a means
to codify those features of the setup prior to $t$ that are relevant to time
evolution after $t$. The question of how $CN$s or the $CN$ -function are
used to predict the outcomes of experiments is addressed through an
assumption. The general interpretative rule is:

\begin{itemize}
\item[{\bf A3.}]  Suppose the $CN$-function of a setup is $\Psi \left(
t\right) $ and at time $t$ one introduces or removes a filter that blocks
out those components of $\Psi $ characterized by a certain property ${\cal P}
$. Suppose further that this modification of the setup has a negligible
effect on the evolution of $\Psi $ after $t$. Then when the $CN$-function is 
$\Psi \left( t\right) $ property ${\cal P}$ will not be detected. The rule
applies to $CN$s in general.
\end{itemize}

\noindent The application of this rule requires a criterion to quantify the
change in $CN$s when setups are modified. In ref. \cite{Caticha98b} the
criterion adopted was to use the Hilbert norm as the means to measure the
distance between $CN$-functions. To justify this Hilbert norm we argue that
there is a uniquely natural notion of distance in the linear space of $CN$%
-functions; it is given by the Hilbert inner product. The argument exploits
the fact that the components out of which setups are built, the filters,
already supply us with a notion of orthogonality. In order to fix the inner
product an additional assumption is needed. It is a form of the principle of
insufficient reason motivated by symmetry \cite{Caticha98c}:

\begin{itemize}
\item[{\bf A4.}]  If there is no reason to prefer one region of
configuration space over another they should be assigned equal a priori
weight.
\end{itemize}

\noindent From the assumptions {\bf A3} and {\bf A4} one can {\em prove}
that the probability of an outcome in an experiment involving a certain
setup is given by the modulus-squared of the corresponding $CN$. This is the
Born postulate.

The final step in the current state of development of the CAQT is a
justification of why the time evolution of the $CN$-functions in addition to
being linear must also be unitary. For this we must assume \cite{Caticha98c}:

\begin{itemize}
\item[{\bf A5.}]  The experimental setups about which we wish to make
predictions involve no loss of information.
\end{itemize}

\noindent Since the objective of {\bf A5} is to specify more precisely what
are the experimental setups we are trying to make predictions about, it is
in effect contributing to define the subject of quantum theory. It may,
therefore, be more appropriate to include {\bf A5} as part of {\bf A1}. In
any case, in these setups entropy must be conserved and this implies unitary
time evolution. The only subtlety in this last step concerns the
identification of the proper entropy; it is the array entropy, not von
Neumann's.

At this point the reader may judge whether the $CN$s introduced in the CAQT
for the sole purpose of representing $and/or$ and which, with rather mild
and natural assumptions, have proved to be useful tools in predicting the
results of experiments deserve or not to be called `amplitudes'. My own
inclination is to {\em define }amplitudes in this way, {\em i.e.}, as tools
for dealing with $and/or$. I regard this definition as being considerably
more {\em useful} than alternative ones because of the light it sheds on the
issue of how amplitudes and wave functions should be interpreted and on
various connections among the usual quantum postulates.

As in all physical theories assumptions are unavoidable. An important issue
is whether the assumptions are sufficiently natural and compelling that one
may not wish to consider abandoning them. A related and somewhat less
important issue is that of deciding how many of these assumptions it is
legitimate to give up and still insist that the resulting theory be called a
``quantum'' theory. Some practitioners of quantum theory would insist that
theories violating assumption {\bf A1} may be ``theories'' but are not
``quantum theories''. Thus, on the question of whether nonlinear quantum
theories are consistent or not I wrote \cite{Caticha98b}: ``Non-linear
variants of quantum mechanics that preserve the notion of amplitudes violate
natural requirements of consistency. The question of whether it is possible
to formulate non-linear versions of quantum mechanics should not be
formulated as a dynamical question about which non-linear terms one is
allowed to add to the Schr\"{o}dinger equation, but rather it should be
phrased as a deeper kinematical question about whether quantum mechanics
should be formulated in terms of mathematical objects other than amplitudes
and wave functions. However, whatever the nature of those mathematical
objects the requirement that they be manipulated in a consistent manner
should be maintained.'' Naturally, the meaning that I attributed then and
now to the words `amplitude' and `wave function' is that of $CN$s and $CN$%
-functions.

We conclude with some comments about the interesting possibility that the
representation assumed in {\bf A2} might not exist. For the purpose of this
discussion it is convenient to classify quantum theories into three classes:

\begin{itemize}
\item[({\em a})]  Theories that satisfy both {\bf A1} and {\bf A2}.

\item[({\em b})]  Theories that satisfy {\bf A1} but do not satisfy {\bf A2}.

\item[({\em c})]  Theories that do not satisfy {\bf A1}.
\end{itemize}

\noindent This covers all possibilities.

The type-({\em a}) theory is conventional quantum theory: its goal is to
predict the outcomes of experiments involving setups built from parts
connected through $and$ and $or$, and since there exists a representation of 
$and/or$ there is no reason why one should not take advantage of it to
construct a {\em linear} quantum theory. Could one, in addition to the
linear theory, also build a nonlinear theory and use it to make predictions
for these same setups? Perhaps, but the predictions of this non-linear
variant would either have to agree with those of the linear theory and be
equivalent to it (probably just a change of variables), or else risk being
inconsistent. Thus, ``nonlinear variants of type-({\em a}) quantum theory
are inconsistent''. Of course, there remains a tiny loophole provided by the
logical possibility that any of assumptions {\bf A3-A5} might not hold.

Theories of type ({\em b}){\em \ }are those which agree with type-({\em a})%
{\em \ }theories as to the subject matter but for which no representation of 
$and/or$ exists. These are, I believe, the theories to which Finkelstein
might have been referring to (he raises no objections against {\bf A1}). One
thing that can be said about these type-({\em b}) theories is that they must
apply to systems that are sufficiently complicated that the linear quantum
theory does not exist. The point is that if the linear quantum theory
existed it would provide us with a representation for $and/or$ in
contradiction with the hypothesis that {\bf A2} is violated. Notice that the
situation here is not that there exists a conceivable linear theory that
must however be discarded on the grounds that it disagrees with experiment.
The situation is more extreme: the linear theory does not exist. The
question then becomes whether the set of type-({\em b}) theories is empty or
not: the cases of a single particle and other simple systems are excluded,
for quantum fields the situation is not so clear.

Finally, type-({\em c}) theories violate {\bf A1 }and could very well be
nonlinear. These theories include all realistic theories that have the more
ambitious goal of providing a direct description of physical reality. Most
proposals for nonlinear theories appear to be of this kind. Type-({\em c})
theories also include those that seek the more modest goal of predicting the
outcomes of experiments involving setups which however are not built from
parts connected through $and$ and $or$. 

My own conclusion is that in the process of achieving consistency for
nonlinear quantum theories so much must be abandoned that by the time the
resulting nonlinear theory is `consistent' it no longer remains `quantum'.  

\noindent {\bf Acknowledgments-} I am indebted to C. Rodr\'{i}guez for
valuable discussions and to J. Finkelstein for stimulating correspondence.

\end{document}